\documentstyle[12pt,epsfig]{article}

\begin{document}
\begin{center}
{{\Large ASPECTS OF THE ELECTROWEAK THEORY} \\
~\\
{\normalsize R. D. Peccei\\
~\\
\it Department of Physics and Astronomy, UCLA \\
\it Los Angeles, CA 90095-1547}}
\end{center}

\section*{Abstract}

These notes contain some of the material I presented at TASI 96 on the
comparison of the standard model with precision electroweak data.  After
a physical accounting of the dominant electroweak radiative corrections,
including the effects of initial state bremsstrahlung, I examine how data
from LEP and the SLC provides a clear test of the standard model.  Apparent
discrepancies in the $R_b$ and $R_c$ ratios, and their physical resolution,
are also examined in some detail.

\section{Introduction}

Having to lecture on ``Aspects of the Electroweak Theory" at TASI '96, a
school whose principal focus was on strings, duality and supersymmetry,
presented a real challenge.  Was there anything about the Standard Model,
I asked myself, which would be both interesting and of use to the highly
theoretical students attending TASI?  In the end, I decided that perhaps two topics fit the bill: precision tests of the electroweak theory and fermion
masses. 

All students have heard the mantra that the standard model is firmly
established by the amazing coincidence of its theoretical predictions
with precision electroweak data, particularly that emanating from the
$e^+e^-$ colliders operating at the $Z$ mass.  Nevertheless, few students
really have a feel for the difficulties involved in establishing this fact
experimentally, or a good physical understanding for the theoretical basis for
this agreement.  For this reason, as a first topic of my lectures, I decided
to talk in some detail about the physics lying behind the confrontation of
precision data with the electroweak theory.

Strictly speaking fermion masses, on the other hand, lie beyond the standard
model.  That is, in the standard model, masses and mixing parameters are input
quantities, not quantities which are determinable by theory.  Nevertheless,
it is clear that ultimately we want to arrive at a theory where these
parameters are predictable.  For this reason, it is interesting to discuss 
what we know at present about these quantities and understand to what extent
this knowledge could eventually be ``predicted" by a deeper theory.  Thus,
it seemed natural to make this the second topic of my lectures.  In
particular, since the origin of fermion masses probably is tied to physics
at the Planck scale, a central question which I tried to answer in my lectures
was what could be inferred from present data about the structure of lepton 
and quark mass matrices at the Planck scale.  Because most of the material in
the second part of my lectures is contained in my recent article with Wang\cite{PW}, I have written up here only material from the first part of my
lectures.

\section{Confronting the Electroweak Theory with Experiment}

The standard model of electroweak interactions is so well known by now\cite
{books} that I will only briefly sketch its principal elements, mostly
to establish a common notation.

\subsection{Elements of the Electroweak Theory}

The Glashow-Salam-Weinberg electroweak theory is based on a $SU(2)\times U(1)$
gauge theory spontaneously broken to $U(1)_{\rm em}$.  As a result of the
symmetry breakdown, the $W^\pm$ and $Z$ gauge bosons acquire a mass.  One gets
the experimentally ``correct" interrelation between the $W$ and $Z$ boson
masses $[M^o_W = M^o_Z\cos\theta^o_W]$ if the agent causing the symmetry
breakdown transforms as an $SU(2)$ doublet\cite{VR}.  Furthermore, if this
agent is a complex scalar doublet Higgs field (or fields) $\Phi$, then the
Yukawa interactions of this field with fermions are a natural source for the
fermion masses and mixing, after the symmetry breakdown.

The $SU(2)\times U(1)$ covariant Higgs kinetic energy term
\begin{equation}
{\cal{L}}_{\rm kin} = -(D_\mu\Phi)^\dagger (D^\mu\Phi)~,
\end{equation}
with
\begin{equation}
D^\mu\Phi = \left[\partial^\mu-ig_o\frac{\tau_a}{2} W_a^\mu -
ig^\prime_o\left(-\frac{1}{2}\right) Y^\mu\right]\Phi~,
\end{equation}
is the source for the weak boson masses once one assumes that the Higgs field
obtains a non-zero VEV:
\begin{equation}
\langle\Phi\rangle = \frac{v_o}{\sqrt{2}}
\left(
\begin{array}{c}
1 \\ 0 
\end{array}
\right)~.
\end{equation}
This VEV provides mass for the charged fields
$W_\pm^\mu = \frac{1}{\sqrt{2}}(W_1^\mu \mp iW_2^\mu)$ and the linear
combination of neutral fields $g_oW_3^\mu-g_o^\prime Y^\mu$.  Specifically,
defining new fields $Z^\mu$ and $A^\mu$ by
\begin{equation}
\left(
\begin{array}{c}
Z^\mu \\ A^\mu
\end{array}
\right) =
\left[
\begin{array}{cc}
\cos\theta_W^o & -\sin\theta_W^o \\
\sin\theta^o_W & \cos\theta_W^o
\end{array}
\right]
\left(
\begin{array}{c}
W_3^\mu \\ Y^\mu
\end{array}
\right)~,
\end{equation}
where $\tan\theta_W^o = g'_o/g_o$, a simple calculation shows that
\begin{equation}
M_W^o = \frac{1}{2} g_ov_o~; ~~~ M_Z^o = \frac{M_W^o}{\cos\theta_W^o}~; ~~~
M_A^o = 0~.
\end{equation}

The Higgs doublet $\Phi$ contains one physical component, the Higgs boson
$H$ and, effectively, one can write
\begin{equation}
\Phi \equiv \frac{1}{\sqrt{2}}
\left(
\begin{array}{c}
v_o + H \\ 0
\end{array}
\right)~.
\end{equation}
The couplings of $H$ to the gauge fields are fixed by the $SU(2)\times U(1)$
symmetry, but its mass $M_H^o$ is arbitrary, being linked to the unknown
Higgs self-interactions.

The interactions of fermions with the gauge fields are also totally specified
by the transformation properties of the fermions under $SU(2)\times U(1)$.
\footnote{Left-handed fermions are $SU(2)$ doublets while right-handed
fermions are singlets.  The $U(1)$ charges of these fermions are essentially 
set by their electromagnetic charge [cf Eq. (10) below].}
These interactions involve the $SU(2)$ and $U(1)$ currents of the fermions
\begin{equation}
J_a^\mu = \sum_i \bar f_i \gamma^\mu(t_a)_if_i~; ~~~
J_Y^\mu = \sum_i \bar f_i \gamma^\mu y_i f_i~,
\end{equation}
where $((t_a)_i,~y_i)$ are the relevant representation matrices for the
$i^{\rm th}$ fermion, coupled to the corresponding gauge fields:
\begin{equation}
{\cal{L}}_{\rm int} = g_o J_a^\mu W_{a\mu} + g'_o J_Y^\mu Y_\mu~.
\end{equation}
Using the Weinberg angle $\theta_W^o$ and the physical fields 
$(W^\pm,Z,A)$ the above can be rewritten as
\begin{eqnarray}
{\cal{L}}_{\rm int} &=& g'_o \cos\theta_W^o
(J_3^\mu + J_Y^\mu) A_\mu + \frac{g_o}{2\sqrt{2}}
\{J_-^\mu W_{+\mu} + J_+^\mu W_{-\mu}\} \nonumber \\
&+& \frac{g_o}{2\cos\theta_W^o}
\left[2(J_3^\mu-\sin^2\theta_W^o(J_3^\mu + J_Y^\mu))\right] Z_\mu~.
\end{eqnarray}

The above formula provides three physical identifications:

\begin{description}
\item{(i)} One recognizes the electromagnetic current
\begin{equation}
J^\mu_{\rm em} = J^\mu_3 + J^\mu_Y
\end{equation}
from its coupling to the photon field $A_\mu$.
\item{(ii)} Similarly, the strength of this coupling identifies the
electric charge $e_o$ as the combination
\begin{equation}
e_o = g'_o\cos\theta_W^o = g_o\sin\theta_W^o~.
\end{equation}
\item{(iii)} Finally, one identifies a neutral current
\begin{equation}
J^\mu_{\rm NC} = 2(J^\mu_3 - \sin^2\theta^o_W J^\mu_{\rm em})
\end{equation}
as the current which couples to the $Z_\mu$ boson.
\end{description}

Using as parameters $e_o$ and $\sin\theta_W^o$ one can write
\begin{equation}
{\cal{L}}_{\rm int} = e_o J^\mu_{\rm em} A_\mu +
\frac{e_o}{2\sqrt{2}\sin\theta_W^o}
\{J_-^\mu W_{+\mu} + J_+^\mu W_{-\mu}\} +
\frac{e_o}{2\cos\theta_W^o\sin\theta_W^o} J^\mu_{\rm NC} Z_\mu~.
\end{equation}
Using the above, it is easy to see that for weak processes where the momentum
or energy transfer is limited $(q^2 \ll M_W^{o^2},~M_Z^{o^2})$ one can
describe these processes by an effective current-current Lagrangian
\begin{eqnarray}
{\cal{L}}^{\rm weak}_{\rm eff} &=& \frac{i}{2!} \int 
{\cal{L}}_{\rm int}  \otimes  {\cal{L}}_{\rm int} \nonumber \\
&\mbox{\raisebox{-1.3ex}{$\stackrel{\textstyle{\simeq}}
{\scriptscriptstyle{q^2 small}}$}}&
\left(\frac{e_o}{2\sqrt{2}\sin\theta^o_W}\right)^2\frac{1}{(M_W^o)^2}
J_+^\mu J_{-\mu} \nonumber \\ 
&+& \frac{1}{2}
\left(\frac{e_o}{2\cos\theta_W^o\sin\theta_W^o}\right)^2 
\frac{1}{(M_Z^o)^2}
J_{\rm NC}^\mu J_{{\rm NC}~\mu}~.
\end{eqnarray}
Thus the Glashow-Salam-Weinberg theory, in this limit reproduces the Fermi
theory, plus some neutral current interactions.  Writing
\begin{equation}
{\cal{L}}^{\rm weak}_{\rm eff} =
\frac{G_F^o}{\sqrt{2}} \left\{J_+^\mu J_{-\mu} + \rho^o
J^\mu_{\rm NC} J_{{\rm NC}~\mu}\right\}
\end{equation}
one identifies the Fermi constant $G^o_F$ as
\begin{equation}
\frac{G_F^o}{\sqrt{2}} = \frac{e_o^2}{8\sin^2\theta_W^o (M^o_W)^2}~.
\end{equation}
The $\rho$-parameter $\rho^o$ gives the strength of the neutral current
interaction relative to that of the charged current interactions and
one sees that
\begin{equation}
\rho^o = \frac{(M_W^o)^2}{(M_Z^o)^2\cos^2\theta_W^o} = 1~,
\end{equation}
where the last equality obtains for doublet Higgs breaking.

The fermions in the theory also interact with the Higgs field.
The most general Yukawa $SU(2)\times U(1)$ invariant interactions between
the left-handed fermion doublets, the right-handed fermion singlets and the
Higgs doublet takes the form

\begin{eqnarray}
{\cal{L}}_{\rm Yukawa} = &-& \Gamma^u_{ij}(\bar u_i\bar d_i)_{\rm L}
\Phi u_{j{\rm R}} - \Gamma^d_{ij}(\bar u_i\bar d_i) \tilde \Phi d_{j{\rm R}} 
\nonumber \\
&-& \Gamma^\ell_{ij}(\bar \nu_i\bar\ell_i)_{\rm L} \tilde\Phi\ell_{j{\rm R}}
+ {\rm h.c.}
\end{eqnarray}
where $i,j$ are family indices and $\tilde\Phi = i\tau_2\Phi^*$.  From the
above it follows that, after symmetry breakdown when $\Phi$ is replaced
by Eq. (6), ${\cal{L}}_{\rm Yukawa}$ 
results in mass terms and Higgs couplings for fermions
of the same charge

\begin{equation}
{\cal{L}}_{\rm eff} = -\left[\bar u_{i{\rm L}}M^u_{ij}u_{j{\rm R}} +
\bar d_{i{\rm L}} M^d_{ij} d_{j{\rm R}} +\bar\ell_{i{\rm L}} M^\ell_{ij}
\ell_{j{\rm R}}\right]
\left(1 + \frac{H}{v_o}\right) 
+{\rm h.c.}~.
\end{equation}
One can diagonalize the above mass matrices $M^f_{ij}$ by a basis change,
leading to a simple effective interaction in which the Higgs field $H$ couples
directly to the fermion masses
\begin{equation}
{\cal{L}}_{\rm eff} = -\sum_i m_i\bar f_i f_i
\left(1 + \frac{H}{v_o} \right)~.
\end{equation}
This basis change does not affect the NC interactions but introduces a 
unitary mixing matrix $V$ in the charged current interactions of quarks---the
Cabibbo-Kobayashi-Maskawa (CKM) matrix.\cite{CKM}

\subsection{Physics at the $Z$ Resonance}

The electroweak theory has received its most challenging tests from the
precision data gathered at the $e^+e^-$ colliders, LEP and SLC, operating
at the $Z$ resonance.  To analyze the important information contained in the
$Z$ line shape, it is not sufficient to consider only the lowest order
electroweak contribution to the process $e^+e^- \to f\bar f$, even taking
the $Z$ width into account in the $Z$ propagator.  The correct calculation
of the $Z$ line shape requires incorporating both electroweak radiative
corrections and purely photonic bremsstrahlung effects, which substantially
alter the resonance peak.

The electroweak radiative corrections for $e^+e^- \to f\bar f$ are complicated
to do in detail.  However, one can estimate their leading effects.  These
involve

\begin{description}
\item{(i)} leading logarithmic contributions of
$O\left(\frac{\alpha}{\pi}\ln M_Z^2/m^2_f\right)$~;
\item{(ii)} non-decoupling contributions of 
$O\left(\frac{\alpha}{\pi}\frac{m_t^2}{M_Z^2}\right)$~.
\end{description}
In these Lectures I will explain the physical origin of these effects and
show that these corrections can be incorporated in an {\bf improved Born
approximation} involving, properly defined, $\sin^2\theta_W$ and
$\rho$ parameters.

To be able to extract from the data the electroweak parameters that one
wants to compare with the Glashow-Salam-Weinberg theory, it is necessary first
to deconvolute from the data the effects of photon bremsstrahlung.  The
dominant effect arises from the bremsstrahlung of photons from the initial
electrons and positrons.  This, so called, initial state bremsstrahlung both
{\bf decreases} the height and {\bf shifts} the location of the resonance
peak.  One can garner the principal features of initial state bremsstrahlung by
summing its leading logarithmic pieces to all orders in $\alpha$ in what amounts
to a QED version of the familiar QCD evolution equation.\cite{MP}  I will begin
by discussing these bremsstrahlung effects.

It is clear physically that if the initial electron, or positron, emits a
bremsstrahlung photon, its energy will be degraded.  Thus to get to the
resonance peak one will need more energy, $\sqrt{s}$, than what one would
need in the absence of bremsstrahlung.  So this effect shifts the resonance
peak to higher $\sqrt{s}$.  Let $e(x;s)$ be the probability density of finding
an electron (or positron) with energy fraction $x$
in the parent electron.  Then one can write formally the cross section corrected
for bremsstrahlung effects as
\begin{equation}
\sigma^{\rm corr}(s) = \int^1_0 dx_1 dx_2 e(x_1;s)~e(x_2;s)~\sigma(x_1x_2s)~,
\end{equation}
which involves the uncorrected cross-section at lower energy
$\sigma(x_1x_2s)$.  By considering the probability of photon emission,
it is easy to write an evolution equation for the electron probability
density\cite{MP} as a function of energy
\begin{equation}
s~\frac{de(x;s)}{ds} = \frac{\alpha}{2\pi} \int^1_x \frac{dy}{y}
P(x/y) e(y;s)
\end{equation}
where the splitting function $P(z)$ is given by\footnote{The + instruction
below serves to remove potentially singular pieces
in the splitting function (cf\cite{Nachtmann})}
\begin{equation}
P(z) = \frac{1+z^2}{(1-z)_+}~.
\end{equation}
A straightforward calculation shows that, to lowest order in $\alpha$ and to log accuracy,
\begin{equation}
e(x;s) = \delta(1-x) + \frac{\alpha}{2\pi} P(x)
\ln s/m_e^2~.
\end{equation}
Hence one has, to this order,
\begin{equation}
\sigma^{\rm corr}(s) = \sigma(s) + \frac{\alpha}{\pi}
\ln s/m_e^2 \int^1_0 dx \frac{1+x^2}{(1-x)_+} \sigma(xs)~.
\end{equation}
Near the $Z$ resonance, as we shall see, the cross section has a Breit-Wigner
form and the integral above gives another logarithmic factor involving the
$Z$ width, $\Gamma_Z$:
\begin{equation}
I\simeq \sigma(s) \ln\left[\frac{(s-M^2_Z)^2+s^2\Gamma^2_Z/M_Z^2}
{s^2}\right]~.
\end{equation}
Hence
\begin{equation}
\sigma^{\rm corr}(s) = \sigma(s)
\left\{1 + \frac{2\alpha}{\pi} \ln s/m_e^2 \ln
\frac{[(s-M^2_Z)^2+s^2\Gamma_Z^2/M^2_Z]^{1/2}}{s}\right\}~.
\end{equation}

This $O(\alpha)$ formula can be generalized to an all order formula by using
the well known fact that\cite{BN} bremsstrahlung logarithms {\bf exponentiate}.
Thus the curly bracket above can be replaced by: $\{\ldots\}\to e^{\{\ldots\}}$.
Defining
\begin{equation}
\beta(s) = \frac{2\alpha}{\pi} \ln s/m_e^2~; ~~~
r(s) = \frac{[(s-M^2_Z)^2+s^2\Gamma^2_Z/M^2_Z]^{1/2}}{s}
\end{equation}
the corrected cross section becomes near resonance\cite{Berends}
\begin{equation}
\sigma^{\rm corr}(s) \simeq \sigma(s) \exp[\beta \ln r] =
\sigma(s) r^\beta~.
\end{equation}
Using the physical values for the $Z$ mass and width one finds that
$\beta(M_Z^2)\simeq 0.11$ and $r(M^2_Z)\simeq 0.027$ and hence
\begin{equation}
r^\beta|_{\rm resonance} \simeq 0.67~.
\end{equation}
Thus, as intimated, the effect of initial state bremsstrahlung leads to a
substantial decrease in the resonance cross-section.  In addition, using
Eq. (29) it is easy to see that these effects lead to a shift in the peak
of the cross section to
\begin{equation}
(\sqrt{s})_{\rm max} = M_Z + \frac{\pi\beta}{8} \Gamma_Z~.
\end{equation}
Numerically this is a shift of over 100 MeV---which is enormous compared to
the few MeV accuracy with which one knows the $Z$ mass.  Better said, to aim
for such an accuracy in the $Z$ mass by analyzing the $Z$ line shape, it is
absolutely crucial to totally understand the ``trivial" photon bremsstrahlung
corrections.  In practice, the full effects of photon bremsstrahlung---including
also non-leading terms---are taken into account by the experimentalists using
dedicated programs like ZFitter\cite{ZFit} and the ``corrected" data is then
compared to the prediction of the electroweak theory.  In the language of
Eq. (29), what is measured is $\sigma^{\rm corr}(s)$ but by knowing the
bremsstrahlung factor $r^\beta$ one can deconvolve from the data the
wanted theoretical cross section $\sigma(s)$.  The effects of initial state
bremsstrahlung are shown pictorially in Fig. 1.

\begin{figure}
\begin{center}
~\epsfig{file=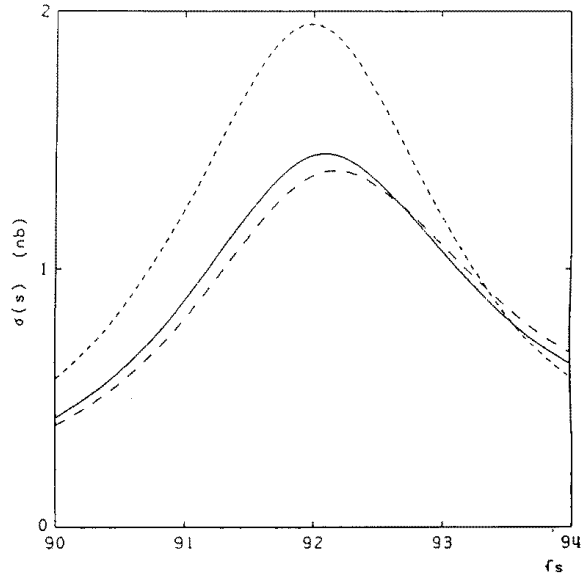,width=8cm}
\end{center}
 \caption{$Z$-line shape: Uncorrected (fine-dashed line); first order
QED correction (dashed line); second order exponential corrections 
(solid line).  From F. Berends[8].}
\end{figure}

Let me now turn to electroweak radiative corrections proper.
The bulk of these corrections for the process $e^+e^- \to f\bar f$
arises from corrections to the gauge propagator.  This is easy to understand since it is only here that different 
mass scales can enter at energy scales $\sqrt{s}
\simeq M_Z$.  The fermion loop in Fig. 2a contains both the scales $M_Z$ and
$m_{f'}$ and the corrections are of 
$O\left(\frac{\alpha}{\pi} \ln M_Z/m_{f'} \right)$.
In contrast, the box graph of Fig. 2b has only $M_W\simeq M_Z$ as the
dominant scale and thus it contributes corrections only of 
$O\left(\frac{\alpha}{\pi}\right)$.  Corrections due to the large top mass of
$O\left(\frac{\alpha}{\pi}\frac{m_t^2}{M_Z^2}\right)$ also arise mostly
through vacuum polarization corrections.  The exception is the process
$e^+e^-\to b\bar b$ which is sensitive to $m_t$ also as a result of the 
vertex graph shown in Fig. 3.

\begin{figure}
\begin{center}
~\epsfig{file=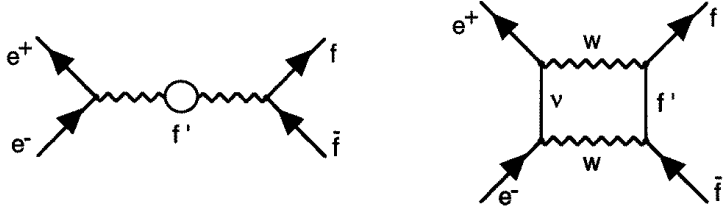,width=10cm}
\end{center}
 \caption{Corrections to the process $e^+e^-\to f\bar f$:
(a) gauge propagator corrections; ~~~ (b) box graph corrections.}
\end{figure}

\begin{figure}
\begin{center}
~\epsfig{file=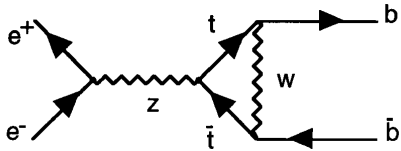,width=6cm}
\end{center}
 \caption{Vertex corrections to the process $e^+e^-\to b\bar b$
sensitive to the top mass.}
\end{figure}

Given the above, it is clear that to understand the principal features of 
electroweak radiative corrections it suffices for our purposes to look at the
corrections to the gauge propagators.  These
corrections are known as {\bf oblique corrections}\cite{oblique}.  A good starting point for
our discussion are the modification to the lowest order photon exchange graph
for the process $e^+e^-\to f\bar f$, shown in Fig. 4a.  This graph leads to
an amplitude

\begin{figure}
\begin{center}
~\epsfig{file=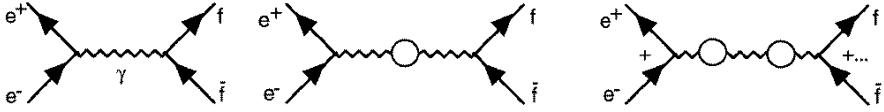,width=12cm}
\end{center}
 \caption[]{(a) Lowest order photon exchange process; 
~~~~(b) Loop modifications to lowest order photon exchange process.}
\end{figure}

\begin{equation}
A^o_\gamma(e^+e^-\to f\bar f) = q_eq_f \frac{e_o^2}{s}
J_{\rm em}^e \cdot J^f_{\rm em}~,
\end{equation}
where $q_e,q_f$ are the numerical values of the fermion electric charges in
units of $e_o$.  The collection of graphs in Fig. 4b produce a modification of
the lowest order photon propagator $1/s$ and changes the bare charge $e_o$
into the physical charge $e$, defined here through its value in Thompson 
scattering at zero momentum transfer [$e=e(o)$].  One has
\begin{equation}
A_\gamma (e^+e^-\to f\bar f) = q_eq_f[e^2\Delta_\gamma(s)]
J^e_{\rm em}\cdot J^f_{\rm em}~.
\end{equation}
In the above, the full propagator $\Delta_\gamma(s)$ contains the self-energy
contributions $\Sigma_\gamma(s) \equiv s\Pi_\gamma(s)$:
\begin{equation}
\Delta^{-1}_\gamma(s) = s+\Sigma_\gamma(s) = s[1+\Pi_\gamma(s)]
\end{equation}
The particular structure shown in Eq. (34) guarantees that indeed the photon
exchange contribution has a pole at $s=0$, as required by gauge invariance.
A comparison of Eq. (33) with Eq. (32) show that for the photon case, the
effect of the radiative corrections is codified through the replacement:
\begin{equation}
\frac{e_o^2}{s}\to \frac{e^2}{s\left[1+\Pi_\gamma(s)\right]} \equiv
\frac{e^2(s)}{s}~,
\end{equation}
with the ``running coupling" $e^2(s)$ defined by
\begin{equation}
e^2(s) = \frac{e^2}{1+\Pi_\gamma(s)}~.
\end{equation}

One can proceed in a similar fashion for the $Z$ exchange contribution.  In this
case, it is useful to write the inverse full propagator as
\begin{equation}
\Delta^{-1}_Z(s) = s-M^2_Z + \Sigma_Z(s) = s-M^2_Z + {\rm Re}~\Sigma_Z(s) +
i~{\rm Im}~\Sigma_Z(s)
\end{equation}
where $M_Z$ is the {\bf physical} mass of the $Z$.  Because $M_Z$ is the
physical mass, to keep the pole location in the $Z$ propagator there, one needs to have
\begin{equation}
{\rm Re} \Sigma_Z (s) = (s-M^2_Z) \Pi_Z(s)~.
\end{equation}
In contrast to the photon case, since the $Z$ has a physical decay channel into
fermions $[Z\to f\bar f]$, one cannot ignore the imaginary part of the $Z$
self-energy.  In fact, denoting schematically the coupling of the $Z$ to fermions by $g_f$, one has that
\begin{equation}
{\rm Im}~ \Sigma_Z (M^2_Z) \sim \sum_f g_f^2~.
\end{equation}

With the above definitions, one can rewrite the inverse $Z$ propagator as
\begin{equation}
\Delta^{-1}_Z(s) = [1 + \Pi_Z(s)] \left\{ s-M^2_Z + i
\frac{{\rm Im}~\Sigma_Z(s)}{[1+\Pi_Z(s)]}\right\}~.
\end{equation}
The last term in the curly bracket above serves to define an $s$-dependent
$Z$ width:
\begin{equation}
M_Z\Gamma_Z(s) =
\frac{{\rm Im}~\Sigma_Z(s)}{[1+\Pi_Z(s)]}~.
\end{equation}
Note that if one defines, analogously to $e^2(s)$, a running coupling of the
$Z$ to fermions via
\begin{equation}
g_f^2(s) = \frac{g^2_f}{[1+\Pi_Z(s)]}~,
\end{equation}
then one sees that the $Z$-width involves precisely the fermion couplings
at this scale
\begin{equation}
\Gamma_Z = \Gamma_Z(M^2_Z) \sim \sum_f g_f^2(M_Z^2)~,
\end{equation}
as one would expect physically.

The radiatively corrected $Z$ contribution, analogous to Eq. (33), reads
\begin{eqnarray}
A_Z(e^+e^-\to f\bar f)&=&g_eg_f[\Delta_Z(s)] J^e_{\rm NC}\cdot
J^f_{\rm NC} \nonumber \\
&=&\frac{g_e(s)g_f(s)}{s-M^2_Z+iM_Z\Gamma_Z(s)}
J^e_{\rm NC}\cdot J^f_{\rm NC}
\end{eqnarray}
One sees that here also the radiative corrections just replace the bare
couplings by the running couplings.  In addition, the $Z$ propagator involves the
physical $Z$ mass and an $s$-dependent width.  Before the bremsstrahlung
corrections, the amplitude in Eq. (44) near the energy corresponding to the 
$Z$ mass leads to a Breit-Wigner formula for
the cross-section for $e^+e^-\to f\bar f$:
\begin{equation}
\sigma^Z_{\rm BW}(s) = \frac{12\pi\Gamma_e\Gamma_f}{M^2_Z}
\left[\frac{s}{(s-M^2_Z)^2+\Gamma^2_Z(s)M^2_Z}\right]~.
\end{equation}
Since, approximately, $\Gamma_Z(s)$ is linear in $s$ near
$\sqrt{s}\simeq M_Z$:
\begin{equation}
\Gamma_Z(s) = \frac{s}{M^2_Z} \Gamma_Z
\end{equation}
it is easy to show that Eq. (45) has a maximum at\footnote{If the $Z$ width
was independent of energy $\Gamma_Z(s) = \Gamma_Z$ the maximum would be
actually shifted in the opposite way to Eq. (47).}
\begin{equation}
(\sqrt{s})_{\rm max} = M_Z - \frac{1}{4}\frac{\Gamma_Z^2}{M_Z} \simeq
M_Z - 17~{\rm MeV}~.
\end{equation}
This small downward shift of the maximum is opposite to the much larger
upward shift caused by initial state bremsstrahlung [cf Eq. (31)].

I anticipate here that the running couplings at $M_Z$ occurring in Eq. (44)
can be written in terms of the Fermi constant measured in $\mu$-decay,
the $Z$ mass and a $\rho$ parameter, which essentially measures the ratio of neutral current
to charged current processes (and which will be defined more precisely shortly):
\begin{equation}
g_e(M^2_Z) g_f(M^2_Z) = \sqrt{2} G_F M_Z^2\rho~.
\end{equation}
This formula is ``sensible" since it just generalizes the lowest order result
\begin{equation}
g^o_eg_f^o = \frac{e_o^2}{4\sin^2\theta_W^o\cos\theta^o_W} = 
\sqrt{2}~G_F^o(M_Z^o)^2\rho^o~,
\end{equation}
but remains to be proven.
Using Eq. (48), one sees that one can incorporate the dominant (logarithmic)
electroweak radiative corrections through an {\bf improved 
Born approximation}\cite{CH}
involving physically measured parameters---including appropriately defined
$\sin^2\theta_W$ and $\rho$ parameters.  One finds
\begin{equation}
A_{\rm Improved~Born} (e^+e^-\to f\bar f) = q_eq_f \frac{e^2(s)}{s} 
J^e_{\rm em}\cdot J^f_{\rm em} +
\frac{\sqrt{2} G_FM^2_Z\ J^e_{\rm NC}\cdot J^f_{\rm NC}}
{s-M^2_Z + iM_Z\Gamma_Z(s)}~,
\end{equation}
where the electromagnetic and neutral currents contain the structures
\begin{equation}
(J^f_{\rm em})_\mu = \gamma_\mu~; ~~~
(J^f_{\rm NC})_\mu = \gamma_\mu\left[t_{3f}(1-\gamma_5)-2q_f\sin^2\theta_W\right]
~.
\end{equation}
In what follows, I elaborate further on the relation between the parameters
$\rho$ and $\sin^2\theta_W$ appearing above and the bare parameters defined
in Section 2.1.

\subsection{Radiative Corrections: Leading Logs}

It is important to understand the relation of the parameters in the improved
Born approximation with the bare parameters which enter in the electroweak
Lagrangian.  The $SU(2)\times SU(1)$ theory has a number of these parameters:
the $SU(2)$ and $U(1)$ couplings, $g_o$ and $g'_o$; the Higgs VEV, $v_o$;
the Yukawa couplings $\Gamma_{ij}^f$; and the Higgs self-coupling, which is
associated with the bare Higgs mass.  These parameters are modified by
radiative corrections.  In fact, when one calculates these corrections one
finds that they are {\bf infinite}.  These infinities must be
eliminated to obtain sensible physical results.  Because the standard model is
a {\bf renormalizable theory} this can be done. It is achieved through a
rescaling of the field appearing in the SM Lagrangian and by replacing the bare
parameters in ${\cal{L}}_{\rm SM}$ by a set of {\bf renormalized} (or
physical) parameters:
\begin{equation}
\{g_o;g_o';v_o,\Gamma^f_{ij},M^o_H\}\to \{g,g',v,(\Gamma^f_{ij})_{\rm ren},
M_H\}~.
\end{equation}
These latter parameters are defined through their relation to specific
measurements.  Because of this, rather than the set of parameters in Eq. (52)
which naturally enter in a Lagrangian description, it is more useful to
choose a more physical set of parameters with which to characterize the theory.

The standard set of parameters which has been adopted in the literature to
describe the electroweak theory replaces the set in Eq. (52) by
\begin{equation}
\{g,g',v,(\Gamma^f_{ij})_{\rm ren},M_H\}\to
\{e,G_F,M_Z,m_{f_i},V_{\rm CKM},M_H\}~.
\end{equation}
Here $m_{f_i}$ are the physical fermion masses defined in terms of zeros
in the inverse fermion propagators (analogous to those we discussed for the
$Z$ boson) and $V_{\rm CKM}$ in the fermion mixing matrix.  For the
most part, when dealing with electroweak precision measurements
at the $Z$ one can 
neglect all fermion masses, except for the top mass, and the effect of fermion
mixing ($V_{\rm CKM}\to 1$).  The electric charge in Eq. (53) is $e(o)$
defined through Thompson scattering, with $\alpha(o) = e^2(o)/4\pi \simeq 1/137$.
$M_Z$ is the pole mass, as defined in Eq. (40), while $G_F$ is the Fermi 
constant as deduced from $\mu$-decay in the Fermi theory, with certain
kinematical QED corrections included\cite{Langacker}.  Having specified a
standard set of parameters, then in the standard model all other measurable quantities
(e.g. the mass of the $W$ boson) are predicted in terms of this standard set.
Physical tests of the electroweak theory are provided by comparing the
prediction for some measurable quantity with its experimentally measured value, e.g.
\begin{equation}
M_W \left|_{\rm theory} = M_W(e,G_F,M_Z,m_t,M_H)\leftrightarrow M_W
\right|_{\rm exp}~.
\end{equation}

It is worthwhile to illustrate the above discussion by
considering in a bit more detail the connection of $G_F$---the Fermi
constant---with $M_W$.  As just indicated, $G_F$ is defined (essentially) as
the effective coupling for $\mu$-decay in the Fermi theory:
\begin{equation}
{\cal{L}}_{\rm eff}^{\rm Fermi} = \frac{G_F}{\sqrt{2}}
(\mu^-\gamma^\mu(1-\gamma_5)\nu_\mu)(\bar\nu_e\gamma_\mu(1-\gamma_5)e)~.
\end{equation}
Using (55) the muon lifetime then provides a direct measure of $G_F$, through
the standard formula $\tau^{-1}_\mu = G_F^2 m^5_\mu/192\pi^3$.  This formula
is modified by pure QED corrections---which are included in the proper 
definition of $G_F$---as these photonic radiative corrections are finite (see below).

If one thinks of Eq. (55) as an effective approximation of the standard model
Lagrangian for values of $q^2\ll M^2_W$, then one would naively expect the
coefficient of the four-fermion operator, $G_F$, to be scale-dependent.
Photon exchange corrections should contribute to the running of $G_F$ 
from the scale $\mu'^2$ to the scale $\mu^2$:
\begin{equation}
G_F(\mu^2) = G_F(\mu'^2)\left[\frac{\alpha(\mu '^2)}{\alpha(\mu^2)}
\right]^{d_\gamma}~,
\end{equation}
with $d_\gamma$ being the anomalous dimension of the 4-Fermi operator in Eq. (55).  However, 
because photon corrections
to the effective Lagrangian (55) give only finite corrections, the anomalous
dimension $d_\gamma$ vanishes.  Thus $G_F(\mu^2)$ is independent of the
scale $\mu^2$: $G_F(\mu^2) \equiv G_F$.  This fact can be easily demonstrated
by a Fierz rotation of Eq. (55), which yields
\begin{equation}
{\cal{L}}_{\rm eff}^{\rm Fermi} = \frac{G_F}{\sqrt{2}}
(\bar\mu\gamma^\mu(1-\gamma_5)e)(\bar\nu_e\gamma_\mu (1-\gamma_5)\nu_\mu)~.
\end{equation}
The second current above, clearly gets no photonic corrections.  So the anomalous dimension $d_\gamma$ is that associated with the $\mu-e$ current.
However, this current
incurs no logarithmic corrections since it is conserved in the limit of neglecting the
$\mu$ and $e$ masses.  Thus, efffectively, $d_\gamma = 0$.

With these considerations, it is easy to convince oneself that to logarithmic
accuracy the effect of radiative corrections is to transcribe directly the
bare relationship between $G_F^o$ and $M_W^o$ into a relationship between
physical quantities.  The bare equation
\begin{equation}
\frac{G_F^o}{\sqrt{2}} = \frac{e_o^2}{8\sin^2\theta_W(M_W^o)^2} =
\frac{e_o^2}{8\left[1-\left(\frac{M_W^o}{M_Z^o}\right)^2\right] (M_W^o)^2}
\end{equation}
becomes, in leading log approximation, a relation between physically
measured quantities.  Provided all 
quantities are measured at the same scale\cite{Marciano}, this equation will
have precisely the same form as the bare equation.  
Thus Eq. (58) is replaced by
\begin{equation}
\frac{G_F}{\sqrt{2}} = \frac{e^2(M_Z^2)}
{8\left[1-\left(\frac{M_W}{M_Z}\right)^2\right]M^2_W}~.
\end{equation}
The RHS of Eq. (59) contains only quantities measured at the scale
of the weak boson masses $M_W \cong M_Z$ and, because $G_F$ is scale 
independent, indeed that is also the scale of
the LHS.  Thus, {\it a fortiori}, Eq. (59) is the correct result!  
One sees that in relating $M_W$ to the standard
set of Eq. (53), all radiative corrections---in leading log approximation---are
contained in the running of the electromagnetic charge from $e^2(o)$ to
$e^2(M^2_Z)$.  Defining, as usual, $\alpha(M^2_Z)=\frac{e^2(M_Z^2)}{4\pi}$
one has the standard model interrelation
\begin{equation}
G_F = \frac{\pi\alpha(M^2_Z)}{\sqrt{2}~M^2_W\left[1-M^2_W/M^2_Z\right]}~,
\end{equation}
which predicts $M_W$ in terms of the standard set of parameters.

In a similar way, also the Weinberg angle can be considered as a derived
quantity in terms of the standard set of parameters.  Recall for these purposes
the different ways in which the Weinberg angle was defined at the Lagrangian
level in terms of bare parameters:
\begin{description}
\item{(i)} Through the unification condition
\begin{equation}
e_o = g_o \sin\theta_W^o = g'_o\cos\theta_W^o~.
\end{equation}
\item{(ii)} Via the $M_W^o-M_Z^o$ relation (doublet Higgs breaking)
\begin{equation}
\sin^2\theta_W^o = 1-\left(\frac{M_W^o}{M_Z^o}\right)^2~.
\end{equation}
\item{(iii)} By comparison to the Fermi theory
\begin{equation}
\frac{G_F^o}{\sqrt{2}} = \frac{e_o^2}{8\sin^2\theta_W^2 (M_W^o)^2}~.
\end{equation}
\item{(iv)} Through the definition of the neutral current, $J^\mu_{\rm NC}$,
\begin{equation}
J^\mu_{\rm NC} = 2(J^\mu_3-\sin^2\theta^o_W J^\mu_{\rm em})~
\end{equation}
\end{description}
One can define different renormalized $\sin^2\theta_W$ by appropriately
generalizing the above definitions.  In general, however, each of these
{\bf renormalized $\sin^2\theta_W$} will be given by a 
slightly {\bf different} function of
the standard set of parameters.

I illustrate the preceding discussion  
by means of two examples.  In the first of these,
the Weinberg angle is given its most physical definition by relating it
directly to the $W$ and $Z$ masses.  This is the definition first adopted by
Sirlin\cite{Sirlin} and one has
\begin{equation}
[\sin^2\theta_W]_s = 1-M^2_W/M^2_Z~.
\end{equation}
From our discussion above of the relation of $G_F$ to $M_W$, it is easy to see
that $[\sin^2\theta_W]_s$ when expressed in terms of the standard set of
electroweak parameters is given (to logarithmic accuracy) by:
\begin{equation}
[\sin^2\theta_W]_s~[\cos^2\theta_W]_s = \frac{\pi\alpha(M^2_Z)}
{\sqrt{2}~G_F M^2_Z}~.
\end{equation}

The second example of a renormalized $\sin^2\theta_W$ which is useful to
consider is the {\bf effective} $\sin^2\theta_W$ which appears in the expression for the neutral current in the improved Born approximation at the
$Z$ resonance:
\begin{equation}
J^\mu_{\rm NC} = 2(J^\mu_3 - \sin^2\theta_{\rm eff} J^\mu_{\rm em})~.
\end{equation}
It is this parameter $\sin^2\theta_{\rm eff}$ which is measured at LEP and the
SLC.  More precisely, $\sin^2\theta_{\rm eff}$ is to be thought of as the
running parameter which multiplies $J^\mu_{\rm em}$ in $J^\mu_{\rm NC}$ at
the scale $\mu^2=M^2_Z$.  That is
\begin{equation}
\sin^2\theta_{\rm eff} = \sin^2\theta_W(M^2_Z)~.
\end{equation}

In the Glashow-Salam-Weinberg model,
the running parameter $\sin^2\theta_W(\mu^2)$ is related to $[\sin^2\theta_W]_s$, with the constant of proportionality being a finite
calculable quantity: 
\begin{equation}
\sin^2\theta_W(\mu^2) = \kappa(\mu^2)[\sin^2\theta_W]_s~.
\end{equation}
A calculation of $\kappa(\mu^2)$\cite{kappa}, again to logarithmic accuracy,
shows that this parameter depends on the scale $\mu^2$ as:
\begin{equation}
\kappa(\mu^2) = 1+\frac{\alpha}{8\pi[\sin^2\theta_W]_s}
\left\{\sum_f (t_{3f}q_f-2q_f^2[\sin^2\theta_W]_s)-1\right\} \ln M^2_Z/\mu^2~.
\end{equation}
Obviously, in view of Eqs. (68-70), in leading log approximation there is
{\bf no difference} between $\sin^2\theta_{\rm eff}$ and $[\sin^2\theta_W]_s$.
However, these quantities differ by the way they depend on $m^2_t$, as will be seen in the next section.

The above discussion serves to justify Eq. (48), which in the improved Born
approximation replaced the product of the effective couplings of the electron
and the outgoing fermion to the $Z$ by the Fermi constant, times the $Z$ mass
squared, times $\rho$.  The coefficient of the effective $Z$ propagator in the
improved Born approximation involves
\begin{equation}
{\rm Coeff} = \frac{e^2}{4[\sin^2\theta_W][\cos^2\theta_W]}
\frac{1}{[1+\Pi_Z(M_Z^2)]}~.
\end{equation}
Using the relation (66) and the definition (36), the above can be rewritten as
\begin{equation}
{\rm Coeff} = \sqrt{2}~G_F M^2_Z\frac{[1+\Pi_\gamma(M^2_Z)]}
{[1+\Pi_Z(M^2_Z)]}~.
\end{equation}
Because in the leading log approximation the only running is that of $\alpha$,
effectively the ratio of the vacuum polarization functions in Eq. (72) is unity.
Furthermore, because $M_W\simeq M_Z$, the ratio of NC/CC contributions also
does not pick up any logarithmic factor.  Hence if one has doublet Higgs
breaking, so that $\rho^o=1$, then in leading log approximation $\rho=1$.  
Whence, to this accuracy, (72) can be written as
\begin{equation}
{\rm Coeff} = \sqrt{2}~G_F M^2_Z\rho~,
\end{equation}
which proves our contention.  However, as we will see, the $\rho$ parameter
does depend on $m^2_t$ and it deviates from unity as a result of
these effects.

\subsection{Radiative Corrections: $m_t$ Effects}

It turns out that the sensitivity to the top mass of physical observables
can all be related to the $m_t$ dependence of the $\rho$ parameter, detailing
the ratio of NC to CC processes.  If one examines the radiative
corrections to the gauge propagators entering in NC and CC processes, it is easy to see that in the Fermi limit, the $\rho$ parameter differs from unity 
only to the
extent that the $Z$ and $W$ self-energies are different from each other
\begin{equation}
\rho = 1+\left[\frac{\Sigma_Z(o)}{M^2_Z}-\frac{\Sigma_W(o)}{M^2_W}\right] =
1 - \left[\Pi_Z(o) - \Pi_W(o)\right]~.
\end{equation}
To compute the vacuum polarization difference in Eq. (74) it suffices to
retain the $t$ and $b$ loops in $\Pi_Z(o)$ and the $t-b$ loop for $\Pi_W(o)$.
Furthermore, since electromagnetic interactions do not give rise to 
contributions proportional to the fermion mass in the loop, it suffices to
retain only the $J^\mu_3$ piece in $J^\mu_{\rm NC}$ for this calculation.  A
simple computation then shows that
\begin{eqnarray}
\Pi_Z(o)-\Pi_W(o) &=& -\frac{3i\alpha}{8\pi^3M^2_W\sin^2\theta_W}\cdot
\int d^4p  \nonumber \\
&\times& \left[\frac{1}{(p^2+m^2_t)^2} + \frac{1}{(p^2+m_b^2)^2}-
\frac{2}{(p^2+m_b^2)(p^2+m_t^2)}\right] \nonumber \\
&=& -\frac{3\alpha}{16\pi\sin^2\theta_W}
\left(\frac{m_t}{M_W}\right)^2~,
\end{eqnarray}
where we have dropped terms of $O(m_b^2)$ relative to those of $O(m_t^2)$.
It follows therefore that
\begin{equation}
\rho = 1+\frac{3G_F}{8\sqrt{2}\pi^2} m_t^2~,
\end{equation}
a formula first obtained by Veltman\cite{Veltman}, which shows that $\rho$ is
quadratically sensitive to the top mass.

To deduce the $m_t$ dependence of other parameters in the theory one can argue
as follows.  The renormalized parameters follow from the bare parameters
by a shift in these parameters.  Thus to trace the $m_t$ dependence it
suffices to track the $m_t$ sensivity of these shifts.  As an example,
consider the Fermi formula, Eq. (16).  The zero order formula, $G_F^o = 
\frac{\pi\alpha^o}{\sqrt{2}~\sin^2\theta_W^o(M_W^o)^2}$, after shifts in
the bare parameters leads to the corrected formula [cf. Eq. (66)]
\begin{equation}
G_F = \frac{\pi\alpha(M^2_Z)}{\sqrt{2}[\sin^2\theta_W]_s[\cos^2\theta_W]_s
M^2_Z} f(m_t^2)~.
\end{equation}
Here the function $f(m_t^2)$ differs from unity only through the $m_t$ 
dependence
arising from the shifts in $\sin^2\theta_W^o$.  
This is because there is no $m_t$
dependence in the shift of the electric charge squared, $\delta\alpha^o$, and
the $m_t$ dependence of $\delta G_F^o$ and $\delta(1/(M_W^o)^2)$ cancel
each other.
Now for the bare quantities one has $\sin^2\theta_W^o =
1-(M_W^o)^2/(M_Z^o)^2$, so that
\begin{equation}
\delta\sin^2\theta_W^o = [\cos^2\theta_W]_s
\left\{\frac{(\delta M_Z^o)^2}{(M_Z^o)^2}-\frac{(\delta M_W^o)^2}{(M_W^o)^2}
\right\}~.
\end{equation}
In the limit $m_t \gg M_W,M_Z$, the mass shifts in Eq. (78) can be related to the
$\rho$ parameter since
\begin{equation}
(\delta M_Z^o)^2 = -\Sigma^o_Z(M_Z^2)\simeq -\Sigma_Z^o(o);~~
(\delta M_W^o)^2=-\Sigma_W^o(M^2_W)\simeq -\Sigma_W^o(o)~,
\end{equation}
so that
\begin{equation}
\delta\sin^2\theta_W^o \simeq [\cos^2\theta_W]_s[\rho-1]~.
\end{equation}
It follows, therefore, that
\begin{equation}
f(m_t^2) = \frac{1}{1+\frac{\delta\sin^2\theta_W^o}{[\sin^2\theta_W]_s}} =
\frac{1}{1+\frac{\cos^2\theta_W}{\sin^2\theta_W}\Delta\rho}~,
\end{equation}
where, using Eq. (76),
\begin{equation}
\Delta\rho = \frac{3G_F}{8\sqrt{2}\pi^2} m_t^2~.
\end{equation}
Thus, including not only leading log contributions but also $m_t^2$ effects,
one can express $[\sin^2\theta_W]_s$ (or equivalently $M_W^2$) in terms of the
standard set of parameters via the formula
\begin{equation}
[\sin^2\theta_W]_s[\cos^2\theta_W]_s =
\frac{\pi\alpha(m^2_Z)}{\sqrt{2}G_FM^2_Z}
\frac{1}{1+\frac{[\cos^2\theta_W]_s\Delta\rho}{[\sin^2\theta_W]_s}}~.
\end{equation}

Similar considerations lead to the following expression for the radiatively
corrected $\sin^2\theta_{\rm eff}$ 
(leading log plus $m_t^2$ corrections):
\begin{equation}
\sin^2\theta_{\rm eff} = [\sin^2\theta_W]_s + \Delta\rho
[\cos^2\theta_W]_s
\end{equation}
which leads to the result
\begin{equation}
\sin^2\theta_{\rm eff} \cos^2\theta_{\rm eff} = 
\frac{\pi\alpha(m^2_Z)}{\sqrt{2} G_F M^2_Z} \frac{1}{1+\Delta\rho}~.
\end{equation}
Since $\sin^2\theta_W\simeq 0.25$, it is clear from Eqs. (83) and (85) that
$[\sin^2\theta_W]_s$ depends more strongly on the value of the top mass than
$\sin^2\theta_{\rm eff}$.

The quantity
\begin{equation}
s_o^2 c_o^2 = \frac{\pi\alpha(m^2_Z)}{\sqrt{2} G_FM^2_Z}
\end{equation}
entering in both Eqs. (83) and (85) is extremely well known experimentally,
with its principal error arising from the error which enters in running
$\alpha$ to the $Z$ mass: $\alpha(M^2_Z) = [129\pm 0.1]^{-1}$\cite{Takeuchi}.  
One finds
\begin{equation}
s_o^2c_o^2 = 0.17755\pm 0.00014
\end{equation}
leading to $s_o^2 = 0.2309\pm 0.0003$.  Unfortunately, it appears difficult
to improve the error on $\alpha(M_Z)$, since it arises principally from errors
in the low energy data on $e^+e^-\to {\it hadrons}$ needed to estimate the
contributions of the hadronic vacuum polarization to the running of
$\alpha$\cite{Takeuchi}.  As we shall see $\pm 0.0003$ is the size of the
experimental error on the value of $\sin^2\theta_{\rm eff}$ determined by
precision electroweak data, so already this error is comparable to the
``standard error" arising due to our imperfect knowledge of $\alpha(M_Z^2)$!

I note that Eqs. (83) and (85) only contain the {\bf dominant} contributions 
of the electroweak radiative corrections.  For detailed comparisons with
precision data one must really use the full formulas, which include both terms
of $O\left(\frac{\alpha}{\pi}\right)$ and corrections that depend on the Higgs mass $M_H$.  These
latter corrections are infinite in the limit as $M_H\to\infty$, since the 
standard model becomes a non-renormalizable theory in this limit.  It was
shown by Veltman\cite{screening} that the sensitivity of the electroweak
corrections to $M_H$ is only logarithmic in lowest order in $\alpha$,
becoming quadratic at $O(\alpha^2)$\cite{VV}.  Detailed calculations give, for
example, for large $M_H$ the formulas\cite{LEPB}
\begin{equation}
\rho = 1 + \frac{3G_F}{8\sqrt{2}\pi^2} m_t^2 -
\frac{3G_FM_Z^2}{4\sqrt{2}\pi^2} s_o^2 \ln M_H/M_Z + \ldots \equiv
1 + \Delta\rho
\end{equation}
\begin{equation}
\sin^2\theta_{\rm eff} = s_o - \frac{c_o^2}{c_o^2-s_o^2}
\left\{\Delta\rho - \frac{G_FM_Z^2}{12\sqrt{2}\pi^2}\ln M_H/M_Z + \ldots
\right\}~.
\end{equation}
One sees from the above that for large $M_H$ the effect of 
having a large top mass is
partly cancelled in $\Delta\rho$.  Furthermore, for large $M_H$ one can
actually change the sign of the correction to $\sin^2\theta_{\rm eff}$, so
that $\sin^2\theta_{\rm eff} > s_o^2$---as seen experimentally.

\subsection{Comparison with Experiment}

A convenient way to compare precision data from LEP{ and SLC (plus low energy
neutrino scattering data and the value of $M_W$ measured at the $p\bar p$
colliders) with theoretical expectations is to refer everything back to
$\sin^2\theta_{\rm eff}$, keeping $m_t$ and $M_H$ as free parameters which are 
then fit to the data.  We do not know anything about $M_H$, except for the LEP
bound that $M_H \geq 66 ~{\rm GeV}$\cite{Higgsbound}.  On the other hand, new
and precise information is being gathered on $m_t$ at the Fermilab Tevatron.  At the time of TASI '96, the value I quoted for $m_t$ was that
given in the 1996 Winter conferences.  This value is now superseded, because
more data has been analyzed by the CDF and DO Collaborations.  The value of
$m_t$ obtained by combining the latest CDF and DO results is\cite{DPF}
\begin{equation}
m_t = (175\pm 6)~{\rm GeV}~,
\end{equation}
which is already amazingly accurate and provide strong constraints on the
theory.

Using the improved Born approximation, Eq. (50), it is straightforward to derive formulas for various measurable quantities.  I summarize some of these 
results below:
\vskip.3cm

{\it Leptonic Width.}
\vskip.3cm
The width for the process $Z^o\to \ell\bar\ell~(\ell=e,\mu,\tau)$, in the
limit where one neglects the lepton mass, is given by
\begin{equation}
\Gamma_\ell = \frac{G_F M_Z^3}{6\pi\sqrt{2}}
[g^2_{V\ell} + g^2_{A\ell}]\left(1+\frac{3\alpha}{4\pi}\right)
\end{equation}
where the last factor is a QED correction accounting for radiative leptonic
decays.  In the above, the vector and axial couplings in the standard model are
universal and are given by
\begin{equation}
g^2_{V\ell} = \frac{1}{4} \rho[1-4\sin^2\theta_{\rm eff}]^2~; ~~~
g^2_{A\ell} = \frac{1}{4} \rho~.
\end{equation}

\vskip.3cm
{\it Forward-Backward Asymmetry.}
\vskip.3cm
The number of fermions $f$ produced in the direction of the incoming electron
$(\theta\leq \pi/2)$ compared to those produced in the backward direction has
a simple form on resonance
\begin{equation}
A^f_{FB} = \frac{\sigma^f(\theta\leq \pi/2)-\sigma^f
(\theta\geq \pi/2)}{\sigma^f(\theta\leq \pi/2)+\sigma^f(\theta\geq\pi/2)}|
_{\rm Res} = \frac{3}{4} {\cal{A}}^e {\cal{A}}^f~,
\end{equation}
where
\begin{equation}
{\cal{A}}^f = \frac{2g_{Vf}g_{Af}}{g^2_{Vf}+g^2_{Af}}~,
\end{equation}
with the vector and axial couplings of the fermion $f$ being defined
analogously to Eq. (92).  One sees that already from $\Gamma_\ell$ and
$A^\ell_{FB}$ it is possible to determine $\rho$ and $\sin^2\theta_{\rm eff}$.
However, there are further (redundant) measurements one can make.
\vskip.3cm

{\it $\tau$-Polarization Asymmetry.}
\vskip.3cm
This asymmetry measures the difference between the cross section for producing
right-polarized taus and left-polarized taus at resonance
\begin{equation}
P_\tau = \frac{\sigma_{\tau_{\rm R}}-\sigma_{\tau_{\rm L}}}
{\sigma_{\tau_{\rm R}}+\sigma_{\tau_{\rm L}}}|_{\rm Res}~.
\end{equation}
This asymmetry can be determined by analyzing the angular distribution of the
$\tau$ decay-products.  The actual value of $P_\tau$ depends in detail on the
production angle $\theta$ of the $\tau^-$, relative to the incoming electron.
After a simple calculation one deduces
\begin{equation}
P_\tau(\cos\theta) = -\frac{{\cal{A}}^\tau(1+\cos^2\theta)+2{\cal{A}}^e
\cos\theta}{1+\cos^2\theta+2{\cal{A}}^\tau{\cal{A}}^e\cos\theta}~.
\end{equation}
By analyzing the dependence of $P_\tau$ on $\cos\theta$ in detail, one can
extract from the data independently ${\cal{A}}^\tau$ and ${\cal{A}}^e$---which,
of course, should coincide in the standard model.
\vskip.3cm

{\it Left-Right Asymmetry.}
\vskip.3cm
This is a quantity that requires an initially polarized $e^-$ beam and can
be measured only at the SLC where beams with a high longitudinal polarization
$(\langle P_e\rangle \simeq 80\%)$ can be produced.  The Left-Right asymmetry
measures the difference in cross-section between beams which are either left-
or right-polarized.
On resonance, one has
\begin{equation}
A_{\rm LR} = \frac{\sigma(e^-_{\rm L})-\sigma(e^-_{\rm R})}
{\sigma(e^-_{\rm L})+\sigma(e^-_{\rm R})} = \langle P_e\rangle
{\cal{A}}^e
\end{equation}
Note that $A_{\rm LR}$, as well as the $\tau$-polarization asymmetry, depend
on a quantity ${\cal{A}}^\ell$  which is approximately
{\bf linear} in $1-4\sin^2\theta_{\rm eff}~[{\cal{A}}^\ell \simeq 2(1-4
\sin^2\theta_{\rm eff})]$, while the Forward-Backward asymmetry
$A^\ell_{\rm FB}$ is quadratic in $1-4\sin^2\theta_{\rm eff}$.  Thus, since
$\sin^2\theta_{\rm eff}\simeq 0.25$, the error on this quantity is under
 much better control in
$A_{\rm LR}$ than in $A^\ell_{FB}$. 

I summarize below the results of LEP and SLC using the 1995 data set.  These
results have been updated for the 1996 Summer conferences, with no major changes
(except in one area to be discussed further below).  Because of this, and
because all the 1995 data is collected together in a
joint publication\cite{LEP95}, I decided
to use this slightly older data set for comparison of experiment with theory.
The data is totally consistent with lepton universality, so I will quote only
the combined result for all three leptons species obtained upon averaging
the four LEP experiments.
The leptonic width and the leptonic Forward-Backward asymmetry at LEP are found
to be\cite{LEP95}
\begin{eqnarray}
\Gamma_\ell &=& (83.93 \pm 0.141)~{\rm MeV} \\
A^\ell_{\rm FB} &=& 0.0172 \pm 0.0012~.
\end{eqnarray}
The result from the $\tau$-Polarization asymmetry from LEP, when one combines
the independent (but consistent) values obtained for ${\cal{A}}^e$ and
${\cal{A}}^\tau$ gives
\begin{equation}
{\cal{A}}^\ell_{\rm LEP} = 0.1406 \pm 0.0057~.
\end{equation}
From the Left-Right asymmetry measured at the SLC one obtains a, somewhat
higher, value for this quantity---although consistent within errors:
\begin{equation}
{\cal{A}}^\ell_{\rm SLC} = 0.1551\pm 0.0040~.
\end{equation}
The value of Eq. (99) for $A^\ell_{\rm FB}$, using lepton universality, 
allows one
to infer another independent determination of ${\cal{A}}^\ell$ and one
finds
\begin{equation}
{\cal{A}}^\ell_{\rm FB} = 0.1514\pm 0.0053~.
\end{equation}
The average of the three measurements (100)-(102) gives, finally,
\begin{equation}
\langle{\cal{A}}^\ell\rangle = 0.1506\pm 0.0028~.
\end{equation}

From the results for $\Gamma_\ell$ and $\langle{\cal{A}}^\ell\rangle$ one can
infer values for $g_{V\ell}$ and $g_{A\ell}$ or equivalently, for $\rho$ and
$\sin^2\theta_{\rm eff}$. Assuming the signs of the standard model, one finds
\begin{equation}
g_{V\ell} = -0.03799\pm 0.00071~; ~~~
g_{A\ell} = -0.50111\pm 0.00041
\end{equation}
and the (leptonic) results
\begin{eqnarray}
\rho &=& 1.0044\pm 0.0012 \\
\sin^2\theta_{\rm eff}|_{\rm leptons} &=& 0.23106 \pm 0.00035~.
\end{eqnarray}
As can be seen from Fig. 5, these leptonic results are in very good agreement
with the Glashow-Salam-Weinberg theory for $60~{\rm GeV} < M_H < 1000~{\rm GeV}$ and $m_t = (180\pm 12)~{\rm GeV}$.

\begin{figure}
\begin{center}
~\epsfig{file=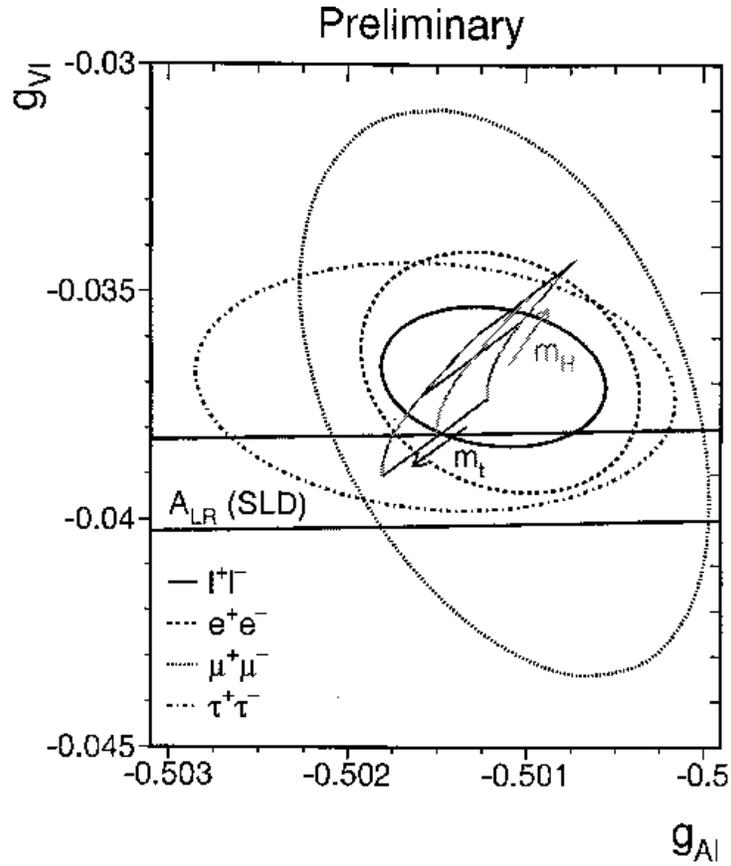,width=10cm}
\end{center}
 \caption[]{$1\sigma$ contours in the $g_{V\ell}-g_{A\ell}$ plane.  The
solid contour assumes lepton universality.  Also shown is the $1\sigma$
band from the $A_{\rm LR}$ measurement.  The grid corresponds to the
standard model prediction for $m_t = 180\pm 12~{\rm GeV}$ and the full Higgs
mass range.  The arrows point in the direction of increasing value of
$m_t$ and $M_H$.  From [23].}
\end{figure}

Further information on $\sin^2\theta_{\rm eff}$ comes from the LEP measurements
of the Forward-Backward asymmetry of heavy quarks ($b$ and $c$).  From these
measurements\cite{LEP95} one can infer values for ${\cal{A}}^b = 0.871\pm
0.029$ and ${\cal{A}}^c = 0.635\pm 0.046$, which imply the (quark) result for
the Weinberg angle:
\begin{equation}
\sin^2\theta_{\rm eff}|_{\rm quarks} = 0.23205\pm 0.00051~.
\end{equation}
The combined result of all measurements, both leptonic and quark, gives
finally a value for $\sin^2\theta_{\rm eff}$ which is accurate to 1 part in
$10^3$:
\begin{equation}
\sin^2\theta_{\rm eff} = 0.23143\pm 0.00028~.
\end{equation}
From this result, in the standard model, one can infer a value for the top mass,
as a function of the Higgs mass.  The value so obtained in\cite{LEP95} is
\begin{equation}
m_t = 180^{+8+17}_{-9-19} {\rm GeV}
\end{equation}
The second error above comes from allowing the Higgs mass to vary from
$M_H = 60~{\rm GeV}$ to $M_H = 1000~{\rm GeV}$.  The central value in
Eq. (109) is that corresponding to assuming $M_H = 300~{\rm GeV}$, with
lower Higgs masses serving to decrease this value.  Obviously, the indirect 
determination of $m_t$ through the standard model radiative corrections is
in excellent agreement with the direct measurement of the top mass at the
Tevatron, Eq. (90).

Individual measurement of various other electroweak quantities are also in
excellent agreement with the standard model\cite{LEP95} except perhaps (at
least for the 1995 data!) for the ratios $R_b$ and $R_c$, 
which measure the ratio of $Z\to b\bar b,~c\bar c$ compared to the total
hadronic rate.  The values that one deduces for $M_W$ and $[sin^2\theta_W]_s$
from the standard model fit, using $m_t = 180~{\rm GeV}$, are also in very
good agreement with direct measurements.  For $M_W$ one predicts\cite{LEP95}
\begin{equation}
M_W|_{\rm SM} = \left(80.359\pm 0.051\pm^{0.023}_{0.024}\right)~{\rm GeV}~,
\end{equation}
to be compared to the average value measured at the collider\cite{Demarteau}
\begin{equation}
M_W = (80.356\pm 0.125)~{\rm GeV}~.
\end{equation}
Similarly, for $[\sin^2\theta_W]_s$ the prediction is\cite{LEP95}
\begin{equation}
[\sin^2\theta_W]_s|_{\rm SM} = 0.2234\pm 0.0009^{+0.0005}_{-0.0012}~,
\end{equation}
while the value one infers from deep inelastic neutrino scattering is
\begin{equation}
[\sin^2\theta_W]_s = 0.2257\pm 0.0047~.
\end{equation}

\subsection{The $R_b-R_c$ Problem and its Resolution}

The 1995 precision electroweak data has one blemish.  The values of the ratios
\begin{equation}
R_b = \frac{\Gamma(Z\to b\bar b)}{\Gamma(Z\to {\rm hadrons})}~; ~~~
R_c \equiv \frac{\Gamma(Z\to c\bar c)}{\Gamma(Z\to {\rm hadrons})}
\end{equation}
appear to be in significant contradiction with the expectations of the
standard model.  The measured numbers are\cite{LEP95}
\begin{equation}
R_b = 0.2219\pm 0.0017~; ~~~
R_c = 0.1543\pm 0.0074~,
\end{equation}
while one expects in the standard model (for $m_t = 180~{\rm GeV}$ and
$M_H \equiv 300~{\rm GeV}$)
\begin{equation}
R_b|_{\rm SM} = 0.2156~; ~~~ R_c|_{\rm SM} = 0.1724~.
\end{equation}

These are difficult measurements and, furthermore, the measurements are
correlated.  Thus, the $3.5\sigma$ discrepancy in $R_b$ and the $2.5\sigma$
discrepancy in $R_c$ are perhaps not so serious.  Indeed, if instead of using
the measured value for $R_c$ in trying to subtract the background in the
$R_b$ measurement, one uses the standard model value for $R_c$ then the
experimental value of $R_b$ changes to
\begin{equation}
R_b = 0.2205\pm 0.0016 ~~~~~~~ [R_c=R_c|_{\rm SM}]~.
\end{equation}

At TASI, I indicated that the attitude towards these results is physicist dependent, with some believing that the source of the possible discrepancy is
due to experimental error in identifying heavy flavor events, while others
take this discrepancy seriously and try to find some new physics phenomena
to account for the measured values of $R_b$ and $R_c$.  As a result of new
data which was presented at the Warsaw International Conference on High Energy
Physics in July and at the DPF Meeting in Minneapolis in August, it now
appears that the 1995 results for $R_b$ and $R_c$ are not to be trusted.  For
instance, the ALEPH Collaboration \cite{PA10}
at LEP and the SLD Collaboration\cite{Weiss} at the
SLC, as a result of new analysis, now find (for $R_c=R_c|_{\rm SM}$)
\begin{eqnarray}
R_b|_{\rm ALEPH} &=& 0.2158\pm 0.0009\pm 0.0011 \\
R_b|_{\rm SLD} &=& 0.2149\pm 0.0033\pm 0.0021~.
\end{eqnarray}
These numbers are in excellent agreement with the standard model expectations.
Similarly, ALEPH\cite{PA16} 
gives a value for $R_c$ which, given its large error, seems
much more compatible with the value expected in the standard model
\begin{equation}
R_c|_{\rm ALEPH} = 0.1683\pm 0.0091~.
\end{equation}

It is interesting to understand the reason for the changes between the 95
and 96 results.  Basically these can be ascribed to a better understanding of the
efficiencies for detecting heavy quarks and the degree of correlation
present when one requires that 2 heavy flavor decays are both detected in one
event.  Although one can measure independently the efficiencies for measuring a
$b$-decay or $c$-decay, the different techniques used all have different
backgrounds that must be considered.
For example, using events where one just tags one $b$-decay,
the quantity $R_b$ can be extracted from the number of tagged events once one
knows the efficiencies $\epsilon_b$ for detecting $b$-decay events and 
$\epsilon_c$ for
excluding $c$-decay events:
\begin{equation}
N_{\rm tag} = 2N_{\rm hadronic}\{\epsilon_bR_b + \epsilon_cR_c\}~.
\end{equation}
Similarly, using events where 2 heavy flavor decays are tagged, to extract
$R_b$ one needs to know in addition the tagging efficiency correlation $C$
for tagging simultaneously two such events:
\begin{equation}
N_{\rm 2~tag} = N_{\rm hadronic} C\{\epsilon_b^2R_b + \epsilon^2_cR_c\}~.
\end{equation}
If $\epsilon_c$ is small (as it is) and the tagging efficiency correlation $C=1$,
then one can determine $R_b$ without having to know precisely the $b$
tagging efficiency $\epsilon_b$.  In this case,
\begin{equation}
R_b = \frac{(N_{\rm tag})^2}{4N_{\rm hadronic} N_{\rm 2~tag}}~; ~~~
\epsilon_b = \frac{2N_{\rm 2~tag}}{N_{\rm tag}}~.
\end{equation}
The new analysis of ALEPH uses the data itself to determine the correlation
efficiency $C$.  This is important, even though $C$ is very near unity, since the
purported accuracy for mesuring $R_b$ is at the percent level.  Hence the
results of the new analysis are more reliable than those reported in 95.

Even though the $R_b-R_c$ crisis is now gone, it might be worthwhile
recalling here some of the theoretical disquisitions which were advanced to
``resolve" this discrepancy, since they illustrate the kind of tight 
constraints that one has as a result of the success, otherwise, of the standard
model in describing data.  If $R_b$ is anomalous but $R_c$ is OK, then from
the precision value of the hadronic width measured at LEP
\begin{equation}
\Gamma_{\rm hadronic} = (1744.8\pm 3.0)~{\rm MeV}
\end{equation}
along with the standard model prediction for this width {\bf before} QCD
corrections, one infers a different value for $\alpha_s(M_Z^2)$ than if there
were no anomaly.  One has
\begin{equation}
\Gamma_{\rm hadronic} = (\Gamma^o_{\rm hadronic})_{\rm SM}
\left(1 + \frac{\alpha_s(M^2_Z)}{\pi}\right) +
(R_b-R_b|_{\rm SM})\Gamma_{\rm hadronic}~.
\end{equation}
If there is no anomaly, from the SM fit one deduces that $\alpha_s(M^2_Z)=
0.125\pm 0.005$.  With the 95 value for $R_b$---taking this anomaly
seriously---instead one obtains\cite{Hagiwara} $\alpha_s(M^2_Z)\simeq 0.104\pm
0.008$.  This latter value was in better agreement with the past
determination of $\alpha_s(M^2_Z)$ from deep inelastic scattering, which
gave a value of $\alpha_s(M^2_Z) = 0.112\pm 0.004$.  However, the most recent
analysis of deep inelastic data now give a value which is about $1\sigma$
higher, which is perfectly compatible with $R_b$ having no anomaly at all!

Many models were put forward which give modifications to $R_b$.  In fact,
since $R_b$ is sensitive to modifications involving the top quark, it is
``natural" that $R_b$ shouuld be more sensitive to new
physics.  Already in the standard model, the $Zb\bar b$ vertex has an additional non-oblique radiative correction due to the presence of the $t$ quark, as 
shown in Fig. 3.  The effective neutral current for the $b$ quark has, as a
result, a modified left-handed coupling:
\begin{equation}
(J^\mu_{\rm NC})_b = \gamma^\mu\left[-
\frac{(1+\epsilon_b)(1-\gamma_5)}{2} + \frac{2}{3}\sin^2\theta_{\rm eff}\right]~,
\end{equation}
where the parameter $\epsilon_b$ depends on the top mass and,
for large $m_t$, is
given by\cite{Akundov}
\begin{equation}
\epsilon_b = -\frac{G_Fm_t^2}{4\pi^2\sqrt{2}}~.
\end{equation}

Bamert {\it et al.}\cite{Bamert} extracted from the 95 data modified chiral
couplings for the $Zb\bar b$ vertex.  Writing
\begin{eqnarray}
g_{b{\rm L}} &=& -(1+\epsilon_b)/2 + \sin^2\theta_{\rm eff}/3 +
\delta g_{b{\rm L}} \nonumber \\
g_{b{\rm R}} &=& \sin^2\theta_{\rm eff}/3 + \delta g_{b{\rm R}}~,
\end{eqnarray}
their analysis found that if $\delta g_{b{\rm R}} \simeq 0$, then
$\delta g_{b{\rm L}} \simeq \epsilon_b/2\simeq -0.0065$.  On the other hand,
if $\delta g_{b{\rm L}}\simeq 0$, then $\delta g_{b{\rm R}}\simeq 
g_{b{\rm R}}/2 \simeq 0.034$.  That is, the effect is reproduced
{\bf either} as a result of a sizeable tree-level right-handed anomalous
coupling {\bf or} as a result of a left-handed anomalous coupling of the size
typical of a loop contribution.

Supersymmetry, as a result of stop-chargino loops, could in principle provide
a $\delta g_{b{\rm L}}$ which is large enough.  Indeed, for light stops and charginos, one finds $\delta g_{b{\rm L}} < 0$ and of the size capable to cancel the
standard model top loop contribution\cite{Kane}.  However, given the limits
on chargino masses from the LEP run at $\sqrt{s} \sim 135~{\rm GeV}$ in Fall
1995, it is difficult to get from supersymmetry an anomalous contribution as large
as $\delta R_b \simeq 0.005$, although half this shift is feasible.  Given
the new trend in the data, this is just as well, since now at best
the $\delta R_b$ needed
is not much larger than 0.002, and could well vanish.

In a similar fashion,\cite{Bamert} it is quite easy to construct a host of
models where through the mixing of the $b$ with another charge -1/3 quark
one generates a rather large $\delta g_{b{\rm R}}$.  Again, before the new
1996 data, Bamert {\it et al.}\cite{Bamert}, as well as others\cite{Rest}, had
identified a number of models which ``fit" the anomalous $R_b$ value.  
With the
disappearance of this anomaly, the interest in these models is moot. 
Nevertheless, since in most of these models the extra parameters
$\delta g_{b{\rm R}}$ and $\delta g_{b{\rm L}}$ were fixed by the 1995 data,
these models remain viable provided one ``turns down" these parameters so as to
agree with the 1996 measurements.  I should note, however, that many of these
models need extra fermions to cancel gauge anomalies.

The 1995 $R_b$ and $R_c$ data also stimulated a number of groups to try to devise
models which would fit {\bf both} the apparently anomalous results (115).  The
simplest consistent new physics 
idea involved imagining that the standard model was
augmented by an extra $U(1)$ symmetry:
\begin{equation}
SU(2)\times U(1) \to SU(2)\times U(1)\times U'(1)
\end{equation}
The 1995
data suggested that this $U(1)'$ had two characteristics:
\begin{description}
\item{(i)} it was generation blind;
\item{(ii)} it was leptophobic.
\end{description}
The first point follows because the sum of the $R_b$ and $R_c$ results of
Eq. (115) is actually considerably below the standard model value.  Thus, if
the extra physics were to act only on the $b$ and $c$ quarks, one needed an
enormous value for $\alpha_s(M_Z^2)$ [$\alpha_s(M_Z^2)=0.185\pm 0.041$\cite{Hagiwara}].  So it made sense to suppose that the new physics
acted on all generations alike, particularly since using the results of Eq.
(115) one found that $3R_b + 2R_c \simeq 1$.

The second requirement above followed because of the good agreement of the
standard model with data on the leptonic sector [cf. Fig. 5].  So if a new
$U(1)'$ existed, it had to have essentially vanishing couplings.That is, this $U(1)'$ was leptophobic and hadrophilic.  One
can then try to determine the universal couplings of this $U(1)'$ to the
up and down quarks of each generation.  The presence of a second $Z$ causes
mixing between the $Z$ and $Z'$ and the size of this mixing, along with the mass
of $Z'$, is restricted by the $\rho$ parameter.  One finds\cite{Alta}
\begin{equation}
\Delta\rho \simeq \xi^2(M^2_{Z'}/M^2_Z)
\end{equation}
where $\xi$ is the mixing angle.  In addition the standard model charges with
which the ordinary $Z$ couples to the quarks are modified because of the
$Z'$ interactions.  One finds
\begin{equation}
\delta q^{\rm NC}_i = \xi[q_{Z'}]_i \tilde g'/g
\end{equation}
where $q_{Z'}$ and $\tilde g'$ are the effective charges and couplings of
the extra $U(1)'$ boson, respectively.

Effectively, these models introduce 5 new parameters: the up and down charges
$q'_{u{\rm R}},q'_{d{\rm R}}$, the coupling $\tilde g'$, and the mass and
mixing of the $Z':M_Z',\xi$.  Not surprisingly,
these models then provide a ``better fit"
than the standard model, even for values of $R_b$ and $R_c$ $3\sigma$ away
from the standard model.  Nevertheless, the up and down couplings determined 
from fitting the existing data were not pleasing.  For instance, Agashe {\it et al.}\cite{Agashe} found, for $M_H = 400~{\rm GeV}$:
\begin{equation}
q'_{u{\rm R}} = 2.19\pm 0.69~; ~~~
q'_{d{\rm R}} = 0.91\pm 0.42
\end{equation}
These hadrophilic couplings are {\bf not} anomaly free for the new $Z'$ and one
must add further quarks in the theory to cancel the $Z'$ anomalies.
Fortunately, the data on $R_b,R_c$ are now much closer to the standard
model, and one does not have to resort to these, frankly ugly, models to
describe the data.

I hope that the discussion of possible extra contributions to $R_b$ and $R_c$
has given the flavor of the difficulty one has to modify only parts of the
standard model and not others, given the already tremendously good fit of
theory with the precision electroweak data.  Thus the new results on $R_b$
(and $R_c$) are a welcome relief, even though one is left with no clues for any
physics beyond the standard model.

\section{Acknowledgments}

This work was supported in part by the Department of Energy under Grant No.
FG03-91ER40662, Task C.


\begin{thebibliography}{99}
\bibitem{PW} R. D. Peccei and K. Wang, Phys. Rev. D{\bf 53} (1996) 2712.
\bibitem{books} For a recent exposition, see for example,
M. E. Peskin and D. V. Schroeder {\bf An Introduction to Quantum Field Theory},
(Addison-Wesley, Reading, MA 1995).
\bibitem{VR} D. A. Ross and M. Veltman, Nucl. Phys. B{\bf 955} (1975) 135.
\bibitem{CKM} N. Cabibbo, Phys. Rev. Lett. {\bf 10} (1963) 531;
M. Kobayashi and T. Maskawa, Prog. Theor. Phys. {\bf 49} (1973) 652.
\bibitem{MP} G. Altarelli and G. Martinelli, in {\bf Physics at LEP},
eds. J. Ellis and R. D. Peccei, CERN Yellow Report 86-02 (1986).  See also,
F. A. Berends, R. Kleiss and S. Jadach, Nucl. Phys. B{\bf 202} (1982) 63.
\bibitem{Nachtmann} See for example, O. Nachtmann, {\bf Elementary
Particle Physics} (Springer Verlag, Berlin 1990).
\bibitem{BN} F. Bloch and A. Nordsiek, Phys. Rev. {\bf 52} (1937) 54.
\bibitem{Berends} For a more complete treatment, see F. Berends, in
{\bf $Z$ Physics at LEP}, eds. G. Altarelli, R. Kleiss and C. Verzegnassi,
CERN Yellow Report 89-08 (1989).
\bibitem{ZFit} Zfitter: D. Bardin {\it et al.}, CERN-TH 6443/92.
\bibitem{oblique} D. C. Kennedy and B. W. Lynn, Nucl. Phys. B{\bf 322} (1989)
1; D. Yu Bardin {\it et al.}, Z. Phys. C{\bf 44} (1989) 493;
W. Hollik, Fortsch. Phys. {\bf 38} (1990) 165; see also M. Consoli and
W. Hollik in {\bf Z. Physics at LEP}, eds. G. Altarelli, R. Kleiss and
C. Verzegnassi, CERN Yellow Report 89-08 (1989).
\bibitem{CH} M. Consoli and W. Hollik in\cite{oblique}.
\bibitem{Langacker} W. J. Marciano and A. Sirlin, Phys. Rev. Lett. {\bf 61}
(1988) 1815.
\bibitem{Marciano} W. J. Marciano, Phys. Rev. D{\bf 20} (1979) 274;
F. Antonelli and L. Maiani, Nucl. Phys. B{\bf 186} (1981) 269.
\bibitem{Sirlin} A. Sirlin, Phys. Rev. D{\bf 22} (1980) 971.
\bibitem{kappa} For a derivation of this formula, see for example, R. D. Peccei
in TASI 88, {\bf Particles, Strings and Supernovae}, eds. A. Jevicki and
C.-I. Tan (World Scientific, Singapore, 1989).
\bibitem{Veltman} M. Veltman, Nucl. Phys. B{\bf 123} (1977) 1989.
\bibitem{Takeuchi} T. Takeuchi in {\bf International Symposium on Vector
Boson Self-Interactions}, eds. U. Baur, S. Errede and T. M\"uller, AIP
Proceedings 350 (AIP Press, Woodbury, NY, 1996).
\bibitem{screening} M. Veltman, Acta Phys. Pol. B{\bf 8} (1977) 475;
M. B. Einhorn and J. Wudka, Phys. Rev. D{\bf 39} (1989) 2758.
\bibitem{VV} J. Van der Bij and M. Veltman, Nucl. Phys. B{\bf 231} (1984) 205.
\bibitem{LEPB} See, for example, F. Jegerlehner, in TASI 90, {\bf Testing
the Standard Model}, eds. M. Cvetic and P. Langacker (World Scientific,
Singapore, 1991).
\bibitem{Higgsbound} L. Roberts, to appear in the Proceedings of the 28th
International Conference on High Energy Physics, Warsaw, Poland.
\bibitem{DPF} B. Winer, to appear in the Proceedings of DPF 96, Minneapolis,
Minnesota.
\bibitem{LEP95} LEP Electroweak Working Group, CERN-PPE 95-172.
\bibitem{Demarteau} M. Demarteau, to appear in the Proceedings of DPF 96,
Minneapolis, Minnesota.
\bibitem{PA10} ALEPH Collaboration: Contribution PA10-015 to the 28th
International Conference on High Energy Physics, Warsaw, Poland.
\bibitem{Weiss} E. Weiss, to appear in the Proceedings of DPF 96, Minneapolis,
Minnesota.
\bibitem{PA16} ALEPH Collaboration: Contribution PA10-016 to the 28th 
International Conference on High Energy Physics, Warsaw, Poland.
\bibitem{Hagiwara} K. Hagiwara, to appear in the Proceedings of the XVII
International Symposium on Lepton and Photon Interactions at High Energy,
Beijing, China (hep-ph/9512425).
\bibitem{Akundov} A. Akhundov, D. Bardin and T. Riemann, Nucl. Phys. B{\bf 276}
(1986) 1.
\bibitem{Bamert} P. Bamert, C. T. Burgess, J. M. Cline, D. London and  
E. Nardi, Phys. Rev. D{\bf 54} (1996) 4275.
\bibitem{Kane} J. Wells, C. Kolda and G. Kane, Phys. Lett. B{\bf 338} (1994)
219.
\bibitem{Rest} E. Ma, hep-ph/9510289; C. V. Chang {\it et al.}, hep-ph/9601326.
\bibitem{Alta} G. Altarelli {\it et al.}, hep-ph/9601324.
\bibitem{Agashe} K. Agashe {\it et al.}, hep-ph/9604266.
\end{thebibliography}
\end{document}